\documentclass[twocolumn,footinbib,showpacs,reprint,aps]{revtex4-1}
\usepackage{color}
\usepackage{amsfonts,amssymb,amsmath,mathrsfs}
\usepackage{amsbsy}
\usepackage{graphicx}
\usepackage{hyperref}

\makeatletter

\begin{document}

\title{Stabilization and control of Majorana bound states with elongated skyrmions}

\author{Utkan G\"ung\"ord\"u}
\email{ugungordu@unl.edu}
\affiliation{Department of Physics and Astronomy and Nebraska Center for Materials and Nanoscience, University of Nebraska, Lincoln, Nebraska 68588, USA}

\author{Shane Sandhoefner}
\affiliation{Department of Physics and Astronomy and Nebraska Center for Materials and Nanoscience, University of Nebraska, Lincoln, Nebraska 68588, USA}

\author{Alexey A. Kovalev}
\affiliation{Department of Physics and Astronomy and Nebraska Center for Materials and Nanoscience, University of Nebraska, Lincoln, Nebraska 68588, USA}

\begin{abstract}
We show that elongated magnetic skyrmions can host Majorana bound states in a proximity-coupled two-dimensional electron gas sandwiched between a chiral magnet and an $s$-wave superconductor. Our proposal requires stable skyrmions with unit topological charge, which can be realized in a wide range of multilayer magnets, and allows quantum information transfer by using standard methods in spintronics via skyrmion motion. We also show how braiding operations can be realized in our proposal.
\end{abstract}

\maketitle

\section{Introduction}
Majorana bound states (MBSs) offer a promising architecture for realization of a topological quantum computer and memory. Such architecture uses non-Abelian anyons to encode and manipulate quantum information \cite{Kitaev2003}. Since Kitaev's  toy model for creating MBSs using the unpaired sites at the ends of a spinless $p$-wave superconducting wire, it has been shown that a conventional $s$-wave superconductor with spin-orbit coupling (SOC) subject to a Zeeman or proximity-induced exchange field can have effective $p$-wave pairing and thus can also support these nonlocal quasiparticles \cite{Fu2008,Oreg2010,Sau2010,Lutchyn2010,Alicea2011,Duckheim2011,Nakosai2013,Klinovaja2013,Kim2015b}. In systems lacking an extrinsic SOC, an effective SOC can also be provided through a nonuniform magnetic texture or field \cite{Alicea2010,Sau2010,Fatin2016}.  Recently, it has been shown that a magnetic texture provided by a skyrmion is suitable for stabilizing  MBSs \cite{Yang2016}.

Control of magnetic textures, such as domain walls, bubbles and skyrmions, is a well-studied subject in spintronics. MBSs bound to these metastable magnetic solitons can be controlled by well-established methods in spintronics. Among these topological magnetic structures, magnetic skyrmions have recently seen a surge of interest since their first experimental observation \cite{Muhlbauer2009a,Yu2010}. The ultralow threshold currents $\sim 10^5$A/m$^2$ required to move skyrmions, and their ability to deform their shape to move around defects, makes skyrmions an attractive alternative to magnetic domain walls in spintronic applications \cite{Nagaosa2013,Iwasaki2013}. Skyrmions can be driven by a wide range of methods such as charge currents \cite{Knoester2014} and gradients of temperature \cite{Kovalev2012,Kovalev2014a,Kong2013} and magnetic field \cite{Yu2016}. Skyrmions have been experimentally driven close to 100m/s velocities using spin polarized charge currents at room temperature \cite{Woo2015}.

Chiral magnets with Dzyaloshinskii-Moriya (DM) interaction \cite{Dzyaloshinsky1958,Moriya1960} prefer skyrmions or antiskyrmions \cite{Gungordu2016a} with unit topological charge, i.e., single winding number and single spin flip from core to outer region in the radial direction. However, skyrmions with a winding number 2 can be stabilized in dipolar magnets \cite{Yu2014} and frustrated magnets \cite{Leonov2015}. In a case with rotational symmetry, only skyrmions with even winding numbers and high odd spin flip numbers can be used to stabilize MBSs in a proximity-coupled conventional $s$-wave superconductor \cite{Yang2016}.


In this paper, we show that elongated skyrmions, which can be stabilized in ordinary chiral magnets  \cite{Gungordu2016a,Hsu2016,Jiang2015,Lin2016,Camosi2017,Hoffmann2017}, can act as an effective ``quantum wire", and under the right conditions realize Kitaev's toy model \cite{Kitaev2001}, locally hosting a pair of Majorana bound states at its ends. It is known that such effective quantum wires can be also formed using a nonuniform magnetic field generated by an array of magnetic tunnel junctions (MTJs) \cite{Fatin2016}. However, this method allows stabilization and manipulation of MBSs only in the region containing active MTJs, which are fixed and cannot be moved. Magnetic skyrmions, on the other hand, remain stable once created and do not require the presence of a fine-tuned, nonuniform external field. Furthermore, they can be manipulated by injecting uniform spin currents or applying field or temperature gradients, which are standard experimental tools widely available in spintronics.

This paper is organized as follows. In Section \ref{sec:model}, we describe the physical setup we propose to realize MBSs and the model we use to describe it. In the following section, we give our numerical results. In Section \ref{sec:braiding}, we describe how to do braiding of MBSs. Finally, Section \ref{sec:conclusion} concludes the paper.

\section{Model}
\label{sec:model}
We consider a 2D electron gas (2DEG) sandwiched between a conventional $s$-wave superconductor and a chiral magnet nanotrack hosting a skyrmion, with a uniform magnetic field applied along the $z$-axis (see Fig.~\ref{fig:skyrmion}).
We remark that in principle in our proposal it is also possible to use a semiconductor wire in regions with ferromagnetic nanotrack rather than a 2DEG \cite{Kim2015b}.
The 2DEG is modelled by the Bogoliubov--de Gennes (BdG) Hamiltonian
\begin{align}
\mathcal H =& \left[\frac{p^2}{2m} -\mu -\frac{\alpha_R}{\hbar} (\boldsymbol e_z \times \boldsymbol p) \cdot {\boldsymbol \sigma} \right]\tau_z + \nonumber \\ &\Delta e^{i \varphi} \tau_+ +   \Delta e^{-i \varphi} \tau_- +   \frac{1}{2}g \mu_B B \sigma_z - J \boldsymbol n \cdot \boldsymbol \sigma
\label{eq:H}
\end{align}
in the Nambu spinor basis $\Psi = (\psi_\uparrow^\dagger, \psi_\downarrow^\dagger,\psi_\downarrow, -\psi_\uparrow)$, where $\psi_\alpha^\dagger$ is the creation operator with spin $\alpha \in \{\uparrow,\downarrow\}$, $\boldsymbol p = -i \hbar \boldsymbol \nabla$, $m$ is the effective electron mass, $\mu$ is the chemical potential,  $\alpha_R$ is the strength of the Rashba SOC, $\Delta e^{i \varphi}$ is the superconducting pairing potential, $J$ and $\boldsymbol n = \boldsymbol n(x,y)$ are the strength and direction of the  proximity induced exchange field due to the presence of the ferromagnet, $\mu_B$ is the Bohr magneton, and $B$ is the strength of the applied uniform magnetic field along the $z$-axis. $\sigma_i$ and $\tau_i$ are Pauli matrices that respectively act on spin and particle-hole subspaces. In what follows, we will take $\varphi = 0$ without loss of generality. We will also focus on the case of $\alpha_R=0$ first.

Realization of Kitaev's nonlocal Majorana fermions requires a ``spinless" system with $p$-wave pairing at the Fermi level. These criteria can be satisfied in an $s$-wave superconductor with spin-orbit coupling (SOC) with an applied magnetic field \cite{Kitaev2001,Alicea2011,Alicea2012,Nakosai2013}. The presence of a nonuniform magnetic texture in Eq.~\ref{eq:H} provides an effective SOC. This can be seen by going into a reference frame in which the effective ``exchange field" $\boldsymbol M \equiv -J \boldsymbol n + g \mu_B  B \boldsymbol e_z/2$ is uniform and aligned with the $z$-axis by making a local gauge transformation $\boldsymbol M \to \hat R \boldsymbol M \equiv M \boldsymbol e_z$. Spatial and temporal dependence of the magnetic texture induces the covariant derivative $\partial_\mu \to \partial_\mu + \hat U \partial_\mu \hat U^\dagger$, where $\hat U = e^{i \sigma_y M_\theta/2} e^{i \sigma_z M_\phi/2}$ is the SU(2) representation (in the spin space) of the real-space rotation matrix $\hat R$ and $M_\phi,M_\theta$ are components of $\boldsymbol M$ in spherical coordinates, resulting in a texture-dependent shift in momentum. In the rotated frame, this gauge potential can be interpreted as the SU(2) vector field,
\begin{align}
\mathcal H = \left[ \frac{(\boldsymbol p - e\boldsymbol A)^2}{2m} + e \phi -\mu \right]\tau_z + \Delta \tau_x + M \sigma_z,
\label{eq:gauge}
\end{align}
where the four-vector potential is determined by the magnetic texture as $\boldsymbol A \equiv i \hbar \hat U \boldsymbol \nabla \hat U^\dagger/e$, $\phi \equiv -i\hbar \hat U \partial_t \hat U^\dagger/e$ . The terms linear in momentum can be interpreted as an effective SOC, which in turn allows the formation of MBSs \cite{Alicea2010,Sau2010,Fatin2016}. For a slowly changing magnetic texture, which we require in order to avoid excitations that can destroy MBSs, spin scalar potential $\phi$ can be neglected.
This leads to a restriction on the maximum velocity of skyrmion motion, $\hbar v_x/R_c^x \ll \Delta$, where $v_x$ is the skyrmion velocity and $R_c^x$ is the skyrmion core radius along the $x$-direction. The adiabaticity assumption further restricts the skyrmion speed. Since we are concerned with MBSs well below the topological gap, we can get a rough estimate for transitions \cite{Drummond2014} by using the Landau-Zener formula \cite{Landau1932,Zener1932}, which yields the condition $J v_x/R_c^x \ll (E_1-E_0)^2/\hbar$, where $E_0$ and $E_1$ are the energies of the ground state and the first excited level.

To estimate the position of MBSs we study the topological gap. For a system with a nonuniform exchange field, the gap is approximately given by \cite{Fatin2016}
\begin{align}
E_g \approx 2 \left[ M -  \sqrt{\left(\mu - \frac{\hbar ^2 (\partial_i \boldsymbol M)^2}{8m M^2}\right)^2  + \Delta^2} \right],
\label{eq:gap}
\end{align}
when the effective exchange field $\boldsymbol M$ is smooth. The linear closing and reopening of the gap as $M$, $\mu$, $\Delta$ vary is indicative of a topological phase transition \cite{Oreg2010,Kim2015b}.
Regions with positive gap ($E_g>0$) are in topological phase, which may host MBSs depending on the geometry of the region \cite{Fatin2016,Yang2016} (see Fig.~\ref{fig:probability}).

The magnetization on the ferromagnet side is described by the free energy $F = \int d^2r \mathcal F$ where the free energy density is given by
\begin{align}
\mathcal F = \frac{A}{2} (\partial_i \boldsymbol n)^2 + (\hat D \boldsymbol e_i)\cdot (\boldsymbol n \times \partial_i \boldsymbol n) - K_u^\text{eff} n_z^2 + \mu_0 M_s H n_z.\label{eq:free}
\end{align}
Here, $A$ is the ferromagnetic exchange strength, $\boldsymbol n$ denotes the direction of the spin density vector, $\hat D$ is the DM tensor \cite{Gungordu2016a}, $K_u^\text{eff} \equiv K_u - \mu_0 M_s^2/2$ is the effective perpendicular easy-axis anisotropy with contributions from magnetocrystalline anistropy and dipolar interactions, and $\mu_0 H$ is the strength of the applied magnetic field along the $z$-axis.
Chiral magnets which can often be described by Eq.~(\ref{eq:free}) can host triangular- and square-lattice of skyrmions (commonly called skyrmion crystal or SkX phase) depending on the strength of the anisotropy and magnetic field \cite{Muhlbauer2009a,Yu2010}. Isolated skyrmions can also be  generated as metastable quasiparticle excitations. In both cases, symmetries of skyrmions reflect the underlying symmetries of the system.  In particular, systems with broken either surface- or bulk-inversion symmetry prefer rotationally symmetric N\'eel (hedgehog-like) or Bloch (vortex-like) type skyrmions, respectively. Additional asymmetries, which can be due to the cutting angle of the sample or applied strain \cite{Gungordu2016a,Koretsune2015,Shibata2015}, can induce interesting deformations, such as elongation of skyrmions along a fixed axis or even skyrmions with negative charge, i.e., antiskyrmions \cite{Gungordu2016a}.

For the proposal described below, it is important that magnetic skyrmions can be driven by spin currents, as well as by gradients of magnetic field, temperature, and stress. One of the advantages of skyrmions over domains walls in spintronic memory device applications is their flexibility, which allows them to deform their shapes to avoid defects. Due to this flexibility, when the expected size of the skyrmion is larger than the width of the racetrack, skyrmions adapt to the presence of the repulsive force due to the edges by becoming elongated. This way of generating elongated skyrmions has the advantage that the axis of elongation can be controlled by moving the skyrmion through sections of the racetrack (see Fig.~\ref{fig:skyrmion}). We confirm elongation of skyrmions due to constrictions with micromagnetic simulations using mumax$^3$ \cite{Vansteenkiste2014}. Mentioned above dynamical properties of skyrmions will be employed in this proposal in order to manipulate MBSs.

\section{Results}
\label{sec:results}
\begin{figure}
	\includegraphics[width=1\columnwidth]{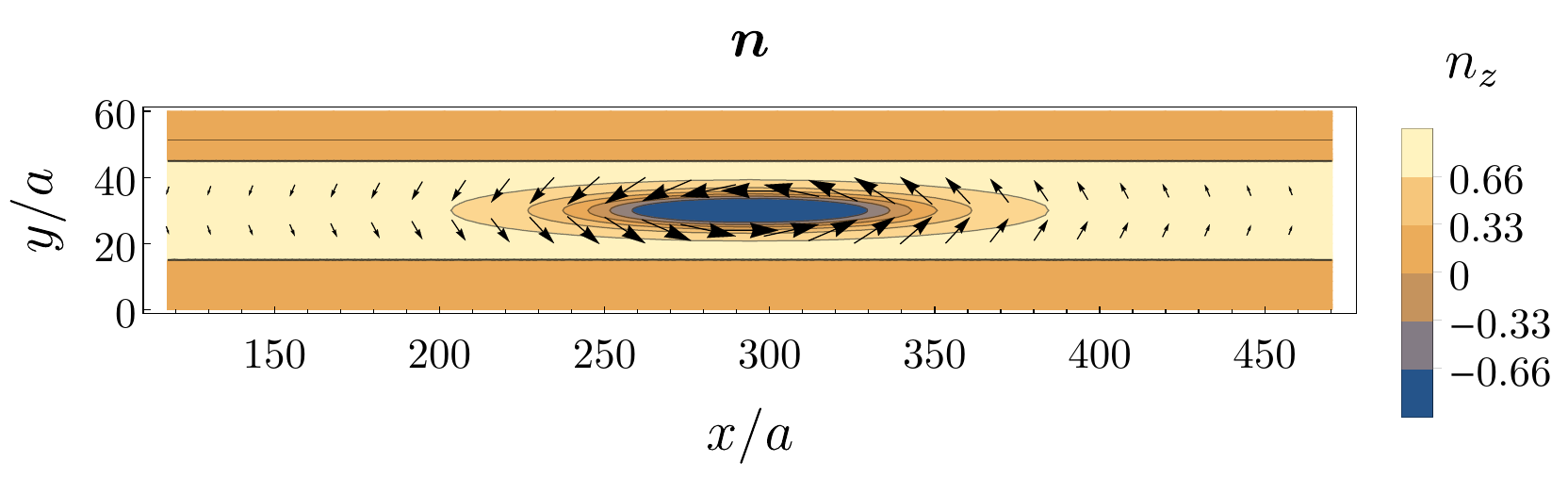}
	\caption{(Color online) Spin density $\boldsymbol n$. Arrows show the in-plane component and contours show the out-of-plane component.}
	\label{fig:skyrmion}
\end{figure}

	For our setup, we consider a skyrmion hosted in the chiral ferromagnetic layer similar to \cite{Yang2016,Pershoguba2016}. We model the magnetic texture with the ansatz $\boldsymbol n = (\sin n_\theta \cos n_\phi, \sin n_\theta \sin n_\phi, \cos n_\theta)$, where the components of the spin density are given by $n_\phi = \phi$, $n_\theta = 2\arctan(R_c^2 /  r^2)$ \cite{Altland2010}, $R_c \sim R/2$ is the core radius where spins become parallel to the plane, and $R$ is the skyrmion radius. We model elongation as stretching of a rotationally symmetric skyrmion, as shown in Fig.~\ref{fig:skyrmion}.
	We ignore the back-action of the superconductor on the chiral magnet, and numerically solve the BdG equation for the eigenenergies and corresponding wavefunctions using the ansatz for a given fixed magnetic texture \footnote{We used Mathematica's NDEigensystem to numerically solve the BdG equation.}.
We will use the following definitions to express the parameters in dimensionless units: $\tilde B\equiv g \mu_B  B/2\Delta$, $\tilde J\equiv J/\Delta$, and $\tilde \mu \equiv \mu/\Delta$.

\begin{figure}
	\includegraphics[width=1\columnwidth]{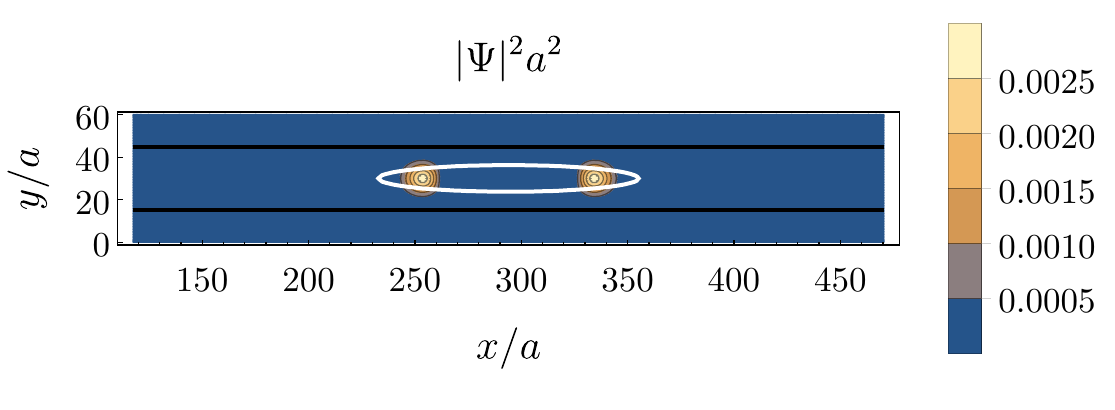}
\caption{(Color online) Squared amplitude $|\Psi|^2$, length in units of $a$ with $a=10$nm for a skyrmion with a core radius of $R_c^y=50$nm, $R_c^x = 490$nm, $\Delta=0.25$meV, $\tilde B=0.87$, $\tilde J=1$, and $\tilde \mu = 0.2$. White solid line is the border between topological and non-topological regions as determined by Eq.~\ref{eq:gap}. Black solid lines indicate the boundaries of the ferromagnetic nanotrack.}
	\label{fig:probability}
\end{figure}

\begin{figure*}
	\includegraphics[width=1.9\columnwidth]{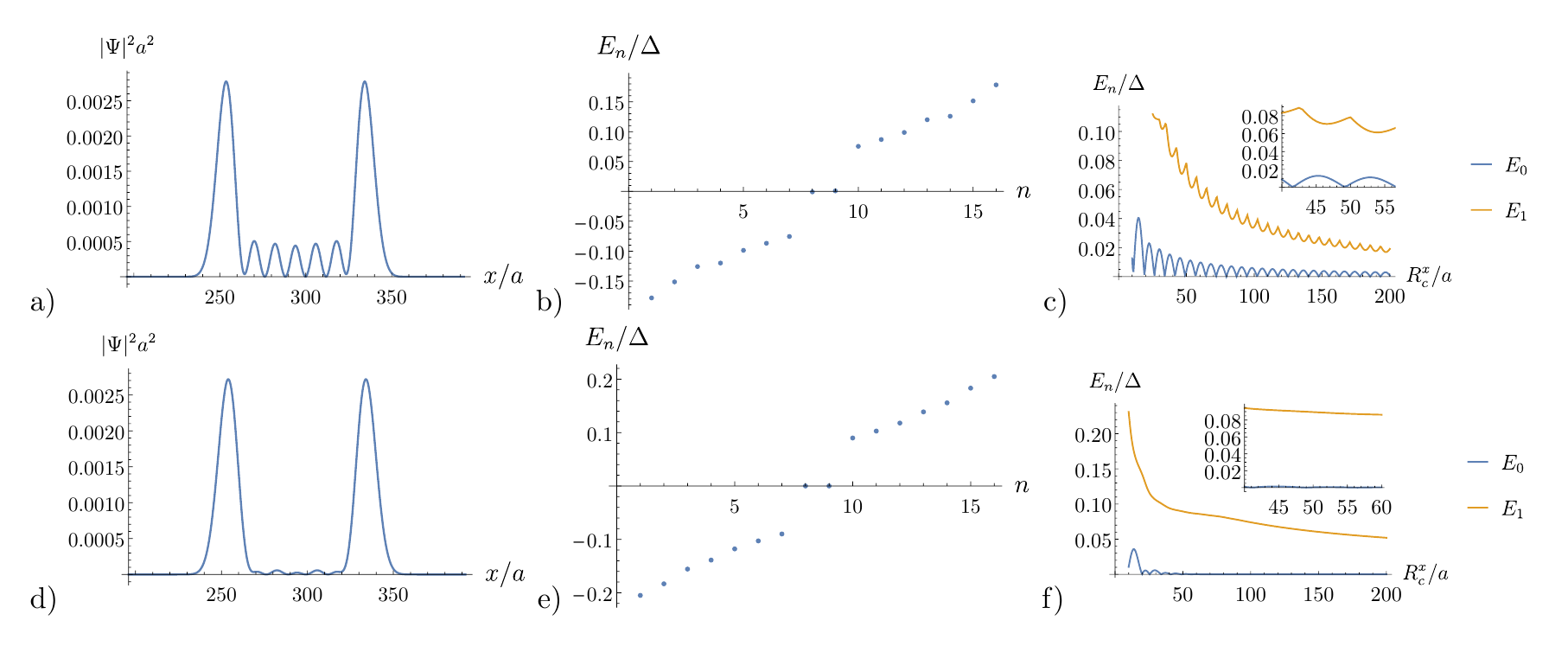}
	\caption{(Color online) a) Squared amplitude $|\Psi|^2$ in units of $1/a^{2}$ along the middle horizontal line for $R_c^x = 490$nm, using parameters given in the caption of Fig.~\ref{fig:probability}. The large peaks correspond to the MBSs localized at the edges of the elliptic topological region. b) Energy spectrum $E_n$ for $R_c^x = 490$nm. c) Ground state ($E_0$) and first excited energy levels ($E_1$) as a function of the horizontal core radius $R_c^x$ in units of $a$. d-f) Similar plots for $\alpha_R = 2.5$meVnm.}
	\label{fig:probabilityx}
\end{figure*}

	A pair of Majorana bound states can be localized at the ends of a topologically non-trivial region which works as an effective quantum wire, as shown in Fig.~\ref{fig:probability} with the white line. There are two cases we consider, one with no extrinsic Rashba SOC and one which includes extrinsic Rashba SOC.
	Figure \ref{fig:probability} shows the squared amplitude for the case with no extrinsic Rashba SOC which has been tuned to achieve Majorana bound states. Figure \ref{fig:probabilityx} shows the squared amplitude and energy spectrum for this case, as well as for the case with extrinsic Rashba SOC. When extrinsic Rashba SOC is included, MBSs have improved localization, and the squared amplitude along the horizontal between the two MBSs flattens considerably as compared to the case with no extrinsic SOC.

 We find that such MBSs can be stabilized over a wide range of elongation, once the strength of the external magnetic field is tuned with regards to the exchange interaction. The spacing between the MBSs is then determined by the amount of elongation.

There is flexibility in parameter tuning for our setup. Figure \ref{fig:probabilityx} shows how the ground state energy level and the first excited energy level change as the elongation is varied. Zero modes are achieved for a skyrmion with a vertical core radius of $R_c^y=50$nm, $\Delta=0.25$meV, $\tilde B=0.87$, $\tilde J=1$ for simplicity, and $\tilde \mu = 0.2$, and $R_c^x$ in the range from $400$nm to $1000$nm. The magnitude of the energy gap $E_1 - E_0$ in units of the superconducting gap $\Delta$ is around 0.08 over this range. Note that the energy levels and the gap can be scaled as $H \to \lambda H$ through replacements $M \to \lambda M$, $\mu \to \lambda\mu$, $\alpha_R \to \sqrt \lambda \alpha_R$, $\Delta \to \lambda \Delta$, $\{x,y\} \to \{x,y\}\sqrt \lambda$, which can be useful in order to find the best material parameters \cite{Fatin2016}.
These values are within reasonable range \cite{Kim2015b,Yang2016,Fatin2016}.

Obtaining MBSs using different values for $\tilde J$ is possible, as long as Eq.~(\ref{eq:gap}) admits closing and reopening of the gap. We remark, however, that using a different value for $\tilde J$ changes the size of the MBSs. This is important because the size of the ellipse-like topological region shown in Fig.~\ref{fig:probability} should be adjusted such that it hosts one and only one mode on each side. A too narrow topological region does not allow MBSs to form, and a too wide topological region allows multiple MBSs which hybridize.
For a given $\tilde J$, the shape of the topological region can be adjusted by choosing a different $R_y^c$ with a constriction of a different size, as well as tuning $\mu$ and $B$.

\begin{figure}
\vspace{1.5em}
\includegraphics[width=0.7\columnwidth]{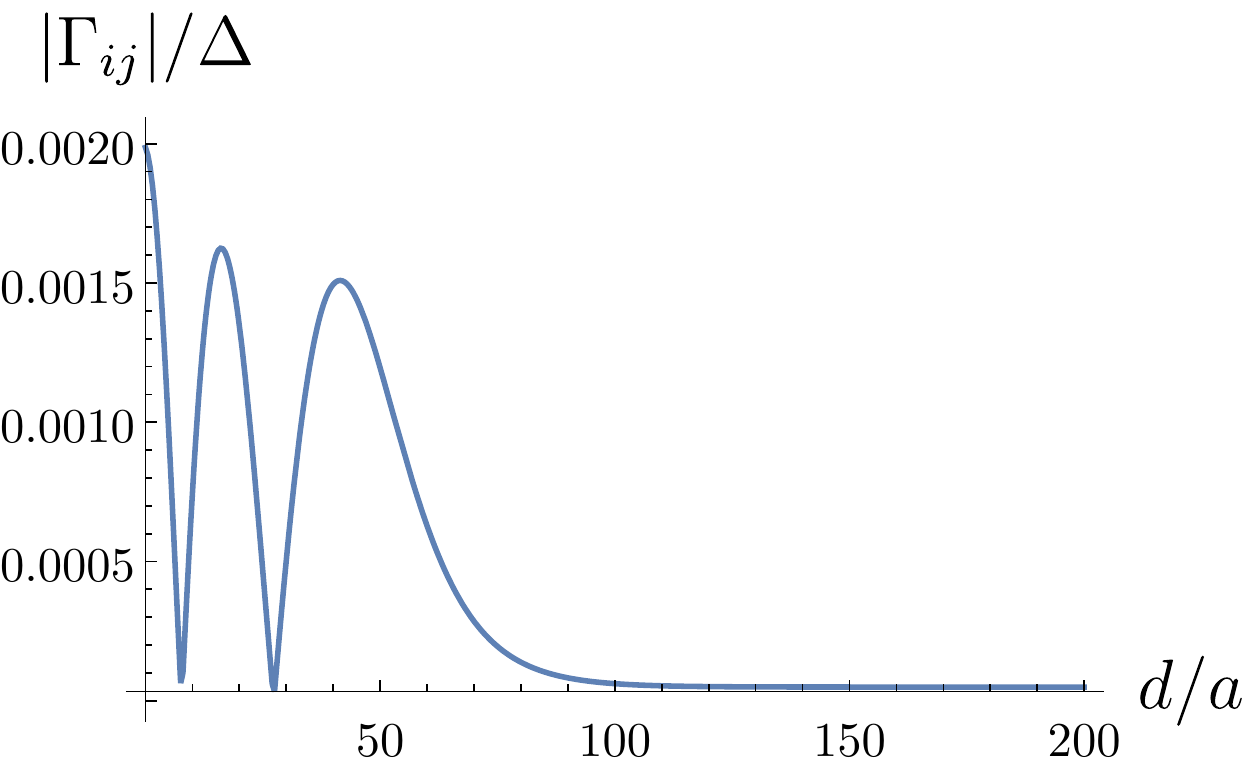}
\caption{(Color online) Hybridization energies $\Gamma_{ij}$ of two MBSs hosted by two different skyrmions in the presence of Rashba SOC, as a function of spatial separation between skyrmions along the $x$-axis, $d$ (skyrmion radius taken to be $R_c^x$).}
\label{fig:gamma}
\end{figure}

In Fig.~\ref{fig:probabilityx}(c), we also plot the energy levels as a function of the distance between MBSs. As the overlap integral between two neighboring MBSs decays exponentially \cite{Drummond2014} (this behavior is also true for MBSs hosted at the ends of two different skyrmions, as shown in Fig.~\ref{fig:gamma}), we observe that the ground state energy becomes small for large $R_c^x$. However, it should be noted that the gap also decays in a similar manner, which makes too large $R_c^x$ undesirable.
As indicated in Eq.~(\ref{eq:gauge}) and the discussion that follows, the gradient of the magnetic texture $\partial_i M$ provides an effective SOC which is required to stabilize MBSs \cite{Alicea2010,Sau2010,Fatin2016}.
The gap's exponential decay observed in Fig.~\ref{fig:probabilityx}(c) is caused by the weakening of the effective SOC which is provided by the texture gradient ($\boldsymbol A \sim \partial_i \boldsymbol n \sim 1/R_c^x$) as the size of the skyrmion increases along the horizontal direction. The presence of an extrinsic SOC stabilizes the gap and improves the localization, which leads to better ground state energetics as shown in the lower plots in Fig.~\ref{fig:probabilityx}.

Figs.~\ref{fig:probabilityx}(c) and (f), taken along with the scaling relations, show the stability of the MBSs for skyrmions with different aspect ratios and sizes. In the absence of an extrinsic SOC, we observe that perturbations in skyrmion size can lead to energetic instabilities. On the other hand, an extrinsic SOC provides a stable operation regime for aspect ratios greater than $\approx10:1$.

\section{Realization of braiding}
\label{sec:braiding}
\subsection{Through Coulomb interaction of Majorana modes}
\begin{figure}
\includegraphics[width=0.5\columnwidth]{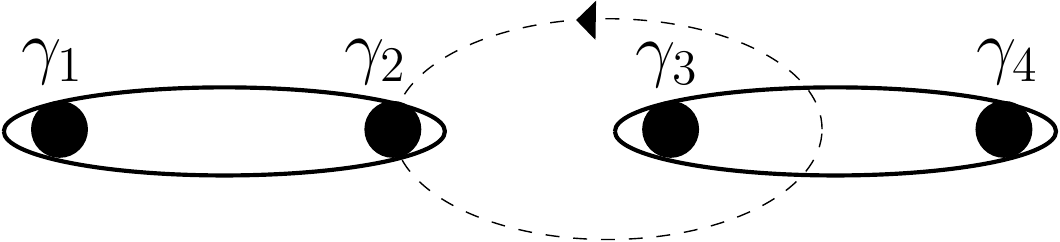}
\caption{A double-braiding of two MBSs at the ends of two topological regions in a typical $p$-wave superconductor in the presence of a vortex, corresponding to two sequential elementary braiding operations given in Eq.~(\ref{eq:braiding}). Note that the topological regions cross each other at intermediate times during the process.}
\label{fig:braiding}
\end{figure}
Realization of a topological quantum computer relies on nonabelian operations through braiding of MBSs. Since implementation of a nontrivial quantum gate requires more than two MBSs, we will discuss braiding operations in a setup with two elongated skyrmions. A typical way of braiding MBSs at the ends of two topological regions involves different regions crossing each other \cite{Alicea2011}, as illustrated in Fig.~\ref{fig:braiding}. However, this is not possible with our ``rigid" regions because such a crossing would involve driving one skyrmion through the other, which would destroy the skyrmions and MBSs. We will instead use an array of Cooper pair boxes, which are superconducting arrays coupled to a large superconductor using a split Josephson junction, such that the magnetic flux through the hole can be used to tune the Josephson energy $E_J$. Such a setup can be used to realize non-abelian braiding operations \cite{Sau2011,VanHeck2012}, and we reproduce the details here for completeness.

\begin{figure}
\includegraphics[width=0.7\columnwidth]{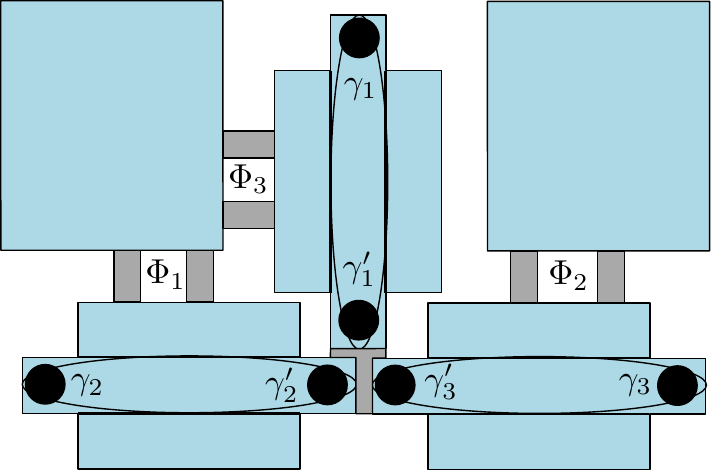}
\caption{Schematic depiction of the three Cooper pair boxes connected at a trijunction. Each Cooper pair box hosts a pair of MBSs and is connected to a bulk superconductor via split Josephson junctions. Coulomb coupling strength can be modulated by changing the applied magnetic flux $\Phi_i$ at each junction.}
\label{fig:trijunction}
\end{figure}

For our proposal, the setup would consist of an elongated skyrmion on each Cooper pair box, with the direction of elongation controlled by the nanotrack on each box. One possible configuration is shown in Fig.~\ref{fig:trijunction}.

The effective low energy Hamiltonian of such a trijunction can be written as \cite{VanHeck2012}
\begin{align}
H_\text{eff} = &i E_M (\gamma_1' \gamma_2' \cos\alpha_{12} + \gamma_2' \gamma_3' \cos\alpha_{23} + \gamma_3' \gamma_1' \cos\alpha_{31}) \nonumber \\
&- i \sum_{k=1}^{3} U_k \gamma_k \gamma_k'
\end{align}
where $E_M$ is the tunnel coupling, $U_k \propto e^{-\sqrt{8 E_J/E_C}}$ is the Coulomb coupling, $E_J = 2 E_0 \cos(\pi \Phi/\Phi_0)$ is the Josephson coupling with $E_0$ as the strength of the coupling, $E_C = e^2/2C$ is the single-electron charging energy with $C$ as the capacitance, $\Phi_0 = h/2e$ is the flux quantum, and the phase differences $\alpha_{ij}$ are given by
$\alpha_{12} = -(\pi/2\Phi_0) (\Phi_1 + \Phi_2 + 2\Phi_3)$,
$\alpha_{23} = (\pi/2\Phi_0) (\Phi_2 + \Phi_3)$,
$\alpha_{31} = (\pi/2\Phi_0) (\Phi_1 + \Phi_3)$.
The Coulomb coupling $U_k \in [U_\text{min}, U_\text{max}]$ decays exponentially with the applied flux, thus $U_\text{max} \gg U_\text{min}$ for the ``on" and ``off" states. It is further assumed that the Coulomb coupling is weaker than the tunnel coupling ($E_M \gg U_k$). As a result, in such a scheme three modes are fused and the four useful modes are $\gamma_1$, $\gamma_2$, $\gamma_3$, and $(\gamma_1^{'}+\gamma_2^{'}+\gamma_3^{'})/\sqrt 3$.

After the flux-controlled Coulomb couplings are turned on and off as depicted in Fig.~\ref{fig:trijunction}, the unitary adiabatic time evolution takes Majorana operators from $\gamma_i$ to $\hat U^\dagger \gamma_i \hat U$ in the Heisenberg picture, where
\begin{align}
\hat U_{ij} = \frac{1+\gamma_i \gamma_j}{\sqrt{2}} + \mathcal O(\epsilon)
\label{eq:braiding}
\end{align}
is the unitary braiding operator \cite{Leijnse2012} and $\epsilon = U_\text{min}/U_\text{max}$ \cite{VanHeck2012}. Since $[\hat U_{ij}, \hat U_{jk}] = \gamma_i \gamma_k$, such operations can be used to realize quantum gates using an array of MBSs.

To braid $\gamma_2$ and $\gamma_3$ in Fig.~\ref{fig:trijunction}, the following sequence needs to be performed: first, to ensure adiabaticity, $\Phi_3$ must be $-\Phi_\text{max}$ at the beginning of the protocol (where $\Phi_\text{max} < \Phi_0/2$); turn $\Phi_1$ to $\Phi_\text{max}$, turn $\Phi_3$ off, turn $\Phi_2$ to $\Phi_\text{max}$, turn $\Phi_1$ off, turn $\Phi_3$ to $-\Phi_\text{max}$, turn $\Phi_2$ off \cite{VanHeck2012}.  During the braiding operation, $\gamma_1$ and $\gamma_1'$ act as ancillary MBSs, which ensure that there is at least one coupling on and one off at each step, such that a two fold degeneracy in the system is maintained.

\subsection{Using measurements}
Aside from the setup which uses Cooper pair boxes described in the previous section, it is also possible to realize measurement-based braiding in different setups. This is achieved by coupling a pair of MBSs to a qubit or a quantum dot. For example, the state of the MBSs can then be projected by measuring the qubit. Various methods for measuring MBSs have been suggested \cite{Vijay2016,Pekker2013,Landau2016,Plugge2016}.

The basic building block of the measurement-based protocol in \cite{Vijay2016} is to perform projective measurements of the operator $i \gamma_i \gamma_j$. Such measurements can be realized in Majorana SQUIDs, MBSs connected with metal bridges forming closed loops \cite{Vijay2016}, either by measuring the persistent current in the SQUID loop through flux measurements, or by measuring the conductance \cite{Vijay2016}.  The operation proceeds only if the results of all measurements are $+1$. A measurement can be described by the operator
\begin{align}
\hat P^{(\pm)}_{\gamma_i \gamma_j} = \frac{1 \pm i\gamma_i \gamma_j}{2}
\end{align}
which acts as identity (null) operator on the $\pm 1$ ($\mp 1$) eigensubspace of $i\gamma_i \gamma_j$.
For example, the following sequence of measurements leads to a nontrivial quantum operation
\begin{align}
\hat P^{(+)}_{\gamma_1^{'} \gamma_1} \hat P^{(+)}_{\gamma_1^{'} \gamma_3^{'}} \hat P^{(+)}_{\gamma_2^{'} \gamma_1^{'}} |\psi\rangle = \frac{1}{2^{3/2}} \hat U_{\gamma_2^{'} \gamma_3^{'}} |\psi\rangle \,,
\label{eq:measurement}
\end{align}
where
\begin{align}
\hat U_{ij} = \frac{1+\gamma_i \gamma_j}{\sqrt{2}}
\label{eq:braiding}
\end{align}
is the unitary braiding operator for $\gamma_i$ and $\gamma_j$. Since $[\hat U_{ij}, \hat U_{jk}] = \gamma_i \gamma_k$, such operations can be used to realize quantum gates using an array of MBSs.

For applications in quantum information, we have to limit ourselves to even or odd parity states. This is due to parity conservation (we neglect quasiparticle poisoning or stray quasiparticle tunneling in and out of the system), i.e., the total fermion number of the system remains even or odd \cite{Saira2012,Riste2013}. Since the operation $\hat U_{ij}$ involves two ``halves" of the two fermions hosted in each skyrmion, it mixes their states.

Since skyrmions can be moved by a variety of methods, we mention that it is possible to move our MBSs by moving the skyrmions which host them. For the set of parameters used for numerical calculations in the previous section, we estimate that the Landau-Zener condition limits the skyrmion velocity as $v_x \ll 1.2$km/s. This is well above the typical velocities for a skyrmion driven by a current or a temperature gradient. In addition, the skyrmion motion should not be too slow as driving skyrmions at $v_x \sim 0.1$m/s over the length of $1 \mu m$ would take $\sim 10\mu$s. This time needs to be well below the decoherence times \cite{Kovalev2014,Drummond2014}.

\section{Conclusion}
\label{sec:conclusion}
We have proposed a way to create Majorana bound states using a conventional $s$-wave superconductor and elongated skyrmions in a typical chiral magnet with Dzyaloshinskii-Moriya interaction. Despite the current lack of experiments coupling magnetic skymions with superconductors, we expect our proposal can be realized in the foreseeable future given that superconductors have been coupled to ferromagnets \cite{Anwar2016}.
A qubit based on such realization should benefit from the topological stability of skyrmions and robustness of quantum operations based on Majorana bound states. Elongated skyrmions can be readily created and manipulated in nanotracks of chiral magnets. While the magnetic texture induced effective SOC is sufficient to realize MBSs, we find that only a setup with extrinsic SOC results in a robust behavior suitable for practical applications, in terms of energy gap and localization of MBSs. Braiding of the MBSs can be realized through the Coulomb interaction of MBSs or a sequence of projective operations.

Disorder may hinder the formation of MBSs by creating additional zero modes localized at random locations. Several possible solutions have been suggested, e.g., using superconductors with weakened disorder, using a tunneling barrier between superconductor and semiconductor, or using a large gap superconductor \cite{Guguchia2011, Adagideli2014, Cole2016}. It might also be necessary to use smaller skyrmions, which should be possible as the size of skyrmions can be tuned over a wide range.

\begin{acknowledgements}
This work was supported by the DOE Early Career Award DE-SC0014189.
\end{acknowledgements}

\bibliographystyle{apsrev}
\bibliography{skyrmion,mbs,extra}

\end{document}